\documentclass{PoS}
\newcommand{\Eq}[1]   {Eq.~(\ref{#1})}

\newcommand{\Fi}[1]   {Fig.~\ref{#1}}

\newcommand{\Ta}[1]   {Table~\ref{#1}}

\newcommand{\gevc}    {\mbox{GeV$/c$}}

\newcommand{\rb}[1]   {\mbox{\textrm{\scriptsize #1}}}

\newcommand{\pt}      {\ensuremath{p_{\rb{T}}}}

\newcommand{\dedx}    {\ensuremath{\textrm{d}E/\textrm{d}x}}

\newcommand{\nwound}  {\ensuremath{\langle N_{\rb{w}} \rangle}}
\newcommand{\sigaver}  {\ensuremath{\langle \sigma \rangle}}

\title{Baryon stopping in 40 and 158 GeV/nucleon Pb+Pb collisions}

\ShortTitle{Baryon stopping in 40 and 158 GeV/nucleon Pb+Pb collisions}

\author{\speaker{H. Str\"obele}\\
        University Frankfurt\\
        E-mail: \email{stroebel@ikf.uni-frankfurt.de}}

\author{for the NA49 collaboration\\}
\abstract{Proton rapidity distributions have been measured by the NA49 collaboration in  40 and 158 GeV/nucleon Pb+Pb collisions as function of collision centrality. We find that the shape and the yield per wounded nucleon in the mid-rapidity region vary little with centrality and are similar to the distributions obtained from inelastic p+p interactions. This observation is satisfactorily described by the transport models HSD and UrQMD, although there are significant differences in the details of the spectral shape 
between the experimental data and the models as well as between the models. The approximate invariance of the normalized proton spectrum in the vicinity of mid-rapidity suggests that multiple nucleon-nucleon interactions in nuclear collisions at SPS energies have little effect on the spectra of those final state protons which are slowed down the most.}

\FullConference{5th International Workshop on Critical Point and Onset of Deconfinement - CPOD 2009,\\
		 June 08 - 12 2009\\
		 Brookhaven National Laboratory, Long Island, New York, USA}

\begin{document}

\section{Introduction}
It is now generally accepted that heavy ion collisions at ultra relativistic energies result in a 
fireball of  matter with high density and temperature. Such conditions prevail, when the 
incoming nucleons deposit enough of their kinetic energy in the reaction zone. Little is known
about this stopping process and about possible differences between the stopping of the incident 
nucleons in elementary nucleon-nucleon interactions and nucleus-nucleus collisions. On a 
microscopic basis and to first approximation each nucleon interacts only once in elementary 
interactions, whereas in central collisions between heavy nuclei it may scatter more than 4 times. The final state distributions of nucleons will be different in the two cases. We study such distributions as function of centrality in Pb+Pb collisions in order to learn about the redistribution process of the incident nucleons. 

Nuclear stopping was studied in central collisions as a function of beam energy at AGS in Brookhaven \cite{Back:2000ru}, at CERN SPS in Geneva \cite{Appelshauser:1998yb} and at RHIC in Brookhaven \cite{Bearden:2003hx}. The inelasticity $K$ is a convenient way to characterize the amount of stopping with a single variable. It is defined as $K  = E_{inel} /(\sqrt{s}/2 -m_p)$  with $E_{inel}$ being equal to the average energy loss of the incident nucleons. 

\begin{figure}[ht]
\vspace{0.8cm}
\begin{minipage}[t]{7cm}
\includegraphics[width=1.0\textwidth]{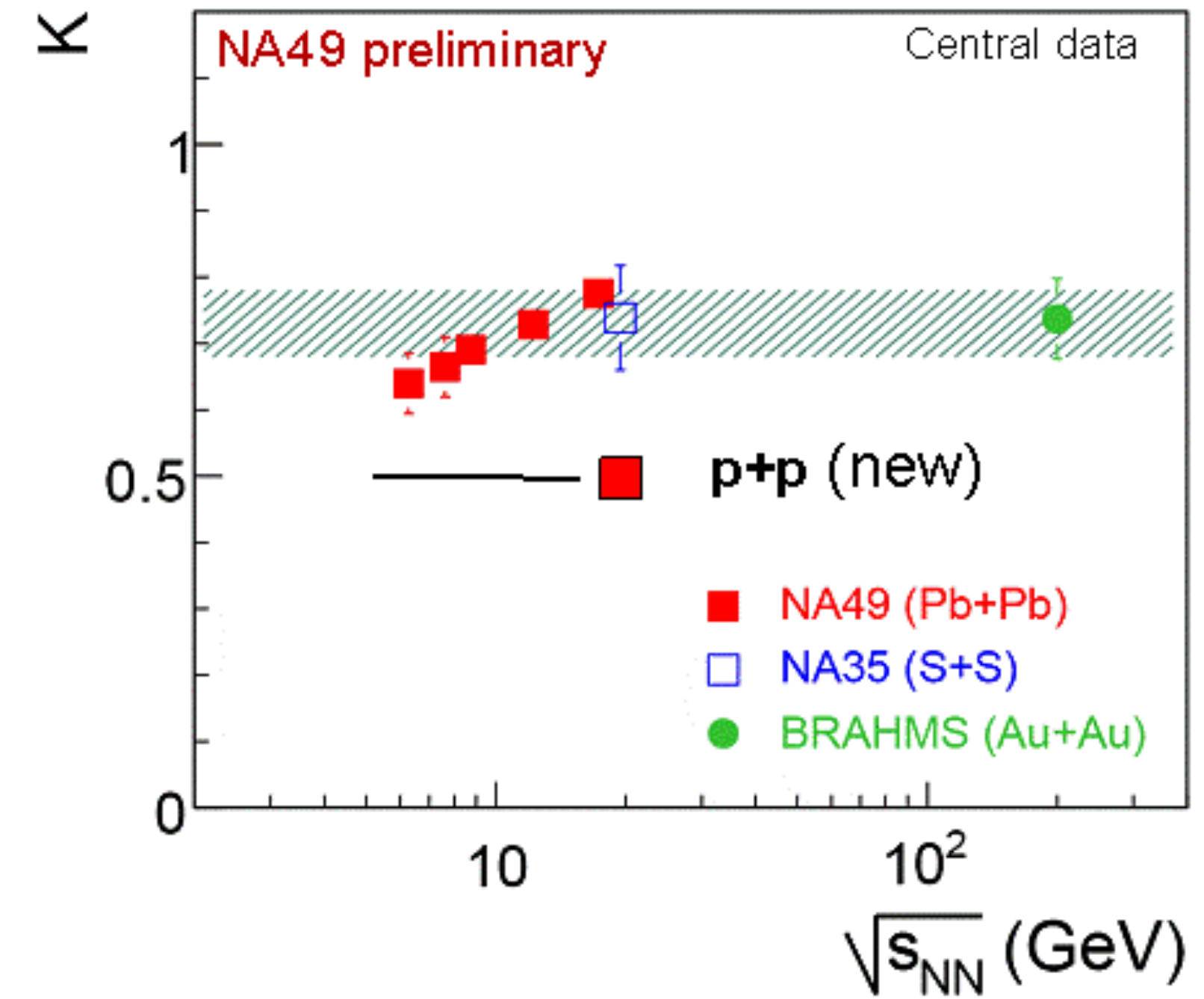}
\caption{The inelasticity $K$ (for the definition see the text) as function of $\sqrt{s_{NN}}$ in central Pb+Pb (NA49), S+S (NA35), and Au+Au (BRAHMS) collisions \protect{\cite{Blume}}. Also shown is the commonly assumed value of $K$ (=0.5)  in inelastic  p+p interactions (black line) and the result from a new NA49 measurement  of inclusive proton spectra in p+p collisions at 158 GeV/c \protect{\cite{Fischer}}.
}
\label{fig:elasticity} 
\end{minipage}
\hspace{0.9cm}
\vspace{-1.0cm}
\begin{minipage}[t]{7cm}
\includegraphics[width=1.0\textwidth]{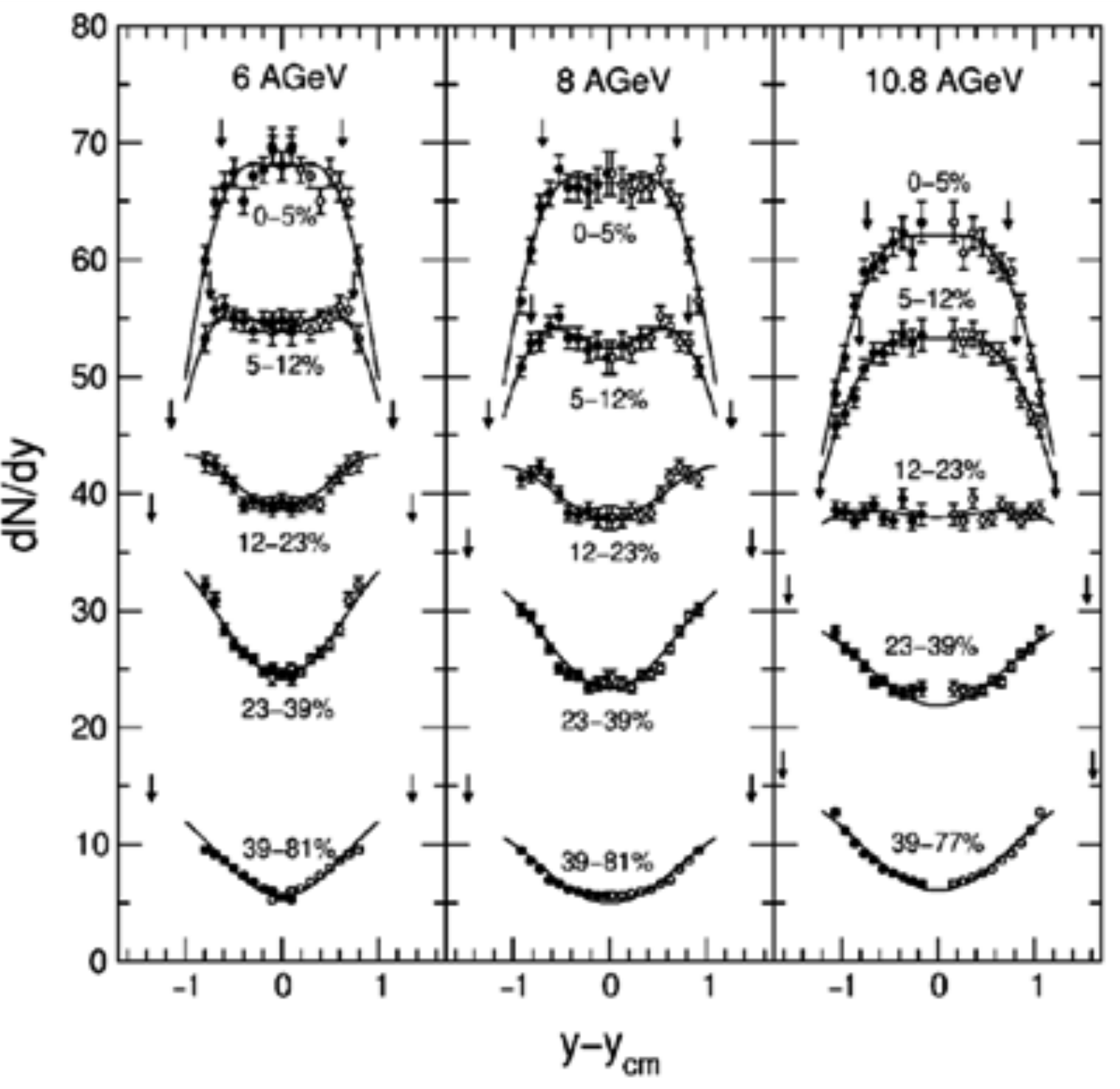}
\caption{
Proton rapidity distributions from Au+Au collisions at 6, 8, and 10.8  GeV/nucleon for different centralities \protect{\cite{Back:2000ru}}.
The arrows indicate the centers of the two Gaussians used to fit the distributions.
}
\label{fig:AGS} 
\end{minipage}
\end{figure} 
\vspace{1.3cm}
\Fi{fig:elasticity} shows $K$ as function of $\sqrt{s}$. The inelasticity in central nuclear collisions is significantly larger than in p+p collisions (K=0.5) as expected \cite{Busza}. In nuclear collisions it increases from 0.65 to 0.78 when comparing AGS with RHIC data \cite{Blume}. This increase goes along with a change of shape of the rapidity distribution, namely a convex form at low (AGS) to a concave one at high (RHIC) energies. The different elasticities of small (p+p) and large (e.g. Pb+Pb) systems should also manifest in a different form of the final state proton rapidity distributions. Such studies have been already performed at AGS by the E917 collaboration at three energies \cite{Back:2000ru}. Their findings are summarized in \Fi{fig:AGS}. Here the rapidity distribution evolves from a concave form in peripheral collisions via a double hump structure (most prominent at 8 GeV/nucleon) to a convex shape in central collisions.

In this contribution we present the centrality dependence of proton rapidity  distributions in Pb+Pb collisions at 40 and 158 GeV/nucleon which were obtained with the NA49 detector.
The phase space coverage extends 1.6 units from mid-rapidity into the forward hemisphere and ranges from zero to 2.0 \gevc~in transverse momentum \pt. 

\section{Experimental setup and data set}

The NA49 detector is a large acceptance hadron spectrometer at the CERN SPS~\cite{NA49NIM}. The main components 
are four large time projection chambers (TPCs) and two super-conducting dipole magnets with a 1 m 
vertical gap, aligned in a row, and a total bending power of 9 Tm. 
Two 2 m long TPCs (VTPCs) inside the magnets allow for precise tracking, momentum determination, 
vertex reconstruction and particle identification (PID) by the measurement of the energy 
loss (\dedx) in the detector gas. The other two TPCs (MTPCs) have large dimensions (4m x 4m x 1.2m) and provide 
additional momentum resolution for high momentum particles as well as PID by \dedx~measurement 
with a resolution of around 4\%.
Two time-of-flight scintillator arrays with close to 1000 photomultipliers each, complement particle identification in the momentum region 3 - 10 GeV/c (around mid-rapidity). A Veto Calorimeter 
(VCAL), which is placed further downstream
along the beam  and covers the projectile spectator phase space region, is used  to select event 
centrality. The NA49-detector is described in detail in \cite{NA49NIM}.
The primary Pb beam had a typical intensity of 10$^{4}$ ions/s and impinged on a target Pb foil
with a thickness of 224 mg/cm$^2$. It passed through several
particle detectors from which the timing signals (resolution 60 - 70 ps) and the individual beam
particle trajectories were determined. A minimum bias trigger was derived from the signal of a gas Cherenkov device right behind the
target. Only interactions which reduce the beam charge as seen by 
this detector by at least 10\% are accepted. The interaction cross 
section thus defined is 5.7 barn. The contamination by fake peripheral events remaining after cuts 
on vertex position and quality amounts to less than 5\% for the most peripheral collisions (see \cite{ALaszlo}). The resulting event ensemble was divided into five centrality bins C0, C1, C2, C3, and C4 
(see \Ta{tab:cent} and \cite{ALaszlo}). The centrality 
selection is based on the forward going energy of projectile spectators as measured in VCAL. The 
number of projectile spectator nucleons and the average number of interacting (wounded) nucleons 
\nwound\  were calculated from the selected cross section fractions using the Glauber approach. 
The track finding efficiency and \dedx\  resolution were optimized by track quality criteria. To 
be accepted, each track must have at least 50 (out of a maximum of 90) potential points in the MTPCs, and 50\% of 
all potential points in the MTPCs must have been reconstructed. In addition, all accepted tracks are required to 
have at least 5 measured and 10 potential points in the VTPCs.

\begin{table}[h]
	\centering
\caption{Cross section fractions 
in \%, average numbers of wounded nucleons \nwound~, numbers of analyzed events and momentum averaged \dedx\ resolutions \sigaver\ for the five centrality classes at
40 and 158 GeV/nucleon.}
\vspace{\baselineskip}
\begin{tabular} {|c||c|c|c|c|c|}
\hline
centrality bin& centrality [ \% ] & \nwound~& accepted events & \sigaver\ [ \% ] \\
%\hline\hline
\hline\hline
\multicolumn {5}{|c|} {40 GeV/nucleon}\\
\hline\hline
C0 & 0-5 &351 $\pm$ 3& 13034& 3.6 $\pm$ 0.2\\
C1 & 5-12.5 &290 $\pm$ 4 & 22971& 3.7 $\pm$ 0.3\\
C2 & 12.5-23.5 & 210 $\pm$ 6 & 34035& 3.7 $\pm$ 0.2\\
C3 & 23.5-33.5&142 $\pm$ 8 &32668& 3.7 $\pm$ 0.2\\
C4 & 33.5-43.5& 93 $\pm$ 7 & 32071& 3.5 $\pm$ 0.2\\
\hline\hline
\multicolumn {5}{|c|} {158 GeV/nucleon}\\
\hline\hline
C0 & 0-5 &352 $\pm$ 3& 15306& 3.8 $\pm$ 0.3\\
C1 & 5-12.5 &281 $\pm$ 4 & 23548& 3.9 $\pm$ 0.2\\
C2 & 12.5-23.5 & 196 $\pm$ 6 & 37053& 3.8 $\pm$ 0.2\\
C3 & 23.5-33.5&128 $\pm$ 8 &34554& 3.8 $\pm$ 0.4\\
C4 & 33.5-43.5& 85 $\pm$ 7 & 34583& 3.7 $\pm$ 0.3\\
\hline
\end{tabular}
\label{tab:cent}
\end{table}

\section{Analysis method}

The protons are identified by the measurement of their specific energy loss \dedx~in the 
relativistic rise region measured in the MTPCs. Their yield was extracted by fitting the function $F(p, \pt)$ (see \Eq{dedx fit}) to the \dedx~distributions of all positively 
(negatively) charged particles in narrow bins of total momentum $p$ and transverse momentum \pt\ . The shape of $F(p,\pt)$ is assumed to be the sum of Gaussians. Their parameters
depend on the particle masses and the measured track lengths. We modified the Gaussian functions by means of an extra asymmetry parameter to account for tails of the Landau distributions which are still present even after truncation of the high \dedx~tails.

The fit function $F$ reads:
\begin{eqnarray}\label{dedx fit}
 F(p,\pt\ ) =	\sum_{i=d,p,K,\pi,e} A_{i} \frac{1}{\sum_{l}n_{l}}\sum_{l}\frac{n_{l}}{\sqrt{2\pi}\sigma_{i,l}} \exp \left[-
\frac{1}{2}{\left(\frac{x-\hat{x_{i}}}{(1\pm\delta) \sigma_{i,l}}\right)}^2\right],
	\end{eqnarray}
where
\begin{itemize}
	\item $A_{i}$ is the raw yield of the particle $i$ under consideration in a given phase space bin,
	\item $n_{l}$ is the number of tracks in a given track length interval $l$. The second sum together
	with the normalization ${\sum_{l}n_{l}}$ forms the weighted average of the track ensembles in each phase space interval,
	\item $\hat{x_{i}}$ is the expected most probable \dedx\ value for particle type $i$ and will be referred to as peak position.
	\item $x$ is the measured \dedx~of a single track,
	\item $\sigma_{i,l}$ is the width for the asymmetric Gaussian of particle type $i$ in length interval $l$, and
	\item $\delta$ is the asymmetry parameter.
\end{itemize}

The amplitude parameters $A_{i}(p,\pt)$ were determined by a maximum likelihood fit to the \dedx~distribution in each $p$, \pt\ bin. The peak positions $\hat{x_{i}}(p)$ are considered to be \pt\ independent and were determined by fits to the \pt\ integrated distributions in each $p$ bin and centrality class. The widths of the Gaussians ($\sigma_{i,l}(p)$) depend on the particle type $i$ and on the track length $l$ according to $\sigma_{i,l} = \sigma\cdot\large(\hat{x_{i}}/{x_{\pi}\large)^\alpha} (1 / \sqrt{l}) $. The exponent $\alpha$ (= 0.625) was extracted from simultaneous fits to $m^2$ distributions from TOF and to \dedx\ distributions from the TPCs \cite{MvL}. The global widths $\sigma(p)$ were determined for each centrality bin by fits to the whole $p$ range . These momentum averaged widths \sigaver\  turned out to be approximately 4 \% for each centrality bin (see \Ta{tab:cent}).
$\delta$ was derived from the observed differences between the widths of the right and left halves of the modified Gaussian functions. This parameter was fixed to be 0.071, since it did not show a significant dependence on centrality, total and transverse momentum. The fit parameters in each $p$, \pt~bin are the five raw yields $A_{i}$ of deuterons, protons, kaons, pions,and positrons. 
These
 yields were transformed from a fine grid in $log(p)$, \pt\ to a coarser grid in $y$,\pt.
 
\subsection{Acceptance and inefficiency}

The raw particle yields have to be corrected for losses due to tracks which do not pass through the detectors or which do not fulfill the
acceptance criteria (acceptance losses) and tracks which are not properly reconstructed (efficiency losses). The acceptance was calculated by generating a sample of protons with flat distributions in transverse momentum and rapidity. The generated particles were propagated through the detectors (and the magnetic field) using the programs provided in the GEANT 3.21 \cite{geant} package. 
Along the resulting trajectories realistic detector signals are generated and processed in exactly 
the same way as the experimental data. The ratio of generated to accepted particle tracks in each 
 $y$,\pt\ bin is the acceptance correction factor.
In addition to the well-defined losses due to the limited and
constrained acceptance, the raw spectra may be subject to losses due to multiplicity and thus centrality dependent inefficiencies. These losses were minimized by restricting the analysis to tracks with azimuthal emission angles within $\pm$ 30 degrees with respect to the bending plane. The remaining losses were determined by the
following procedure. Ten GEANT generated proton tracks and
their signals were embedded into raw data of real events. Only those tracks are embedded which pass all acceptance
criteria. These modified events are reconstructed with the standard
reconstruction chain. The ratio of all generated tracks to those
reconstructed constitutes the correction factor for reconstruction inefficiencies. The resulting efficiencies vary with centrality by less than 5\% and are over 95\% in most $y$,\pt\ bins.

\section{Results}
\begin{center}
\begin{figure}[h]
\includegraphics[width=0.5\linewidth]{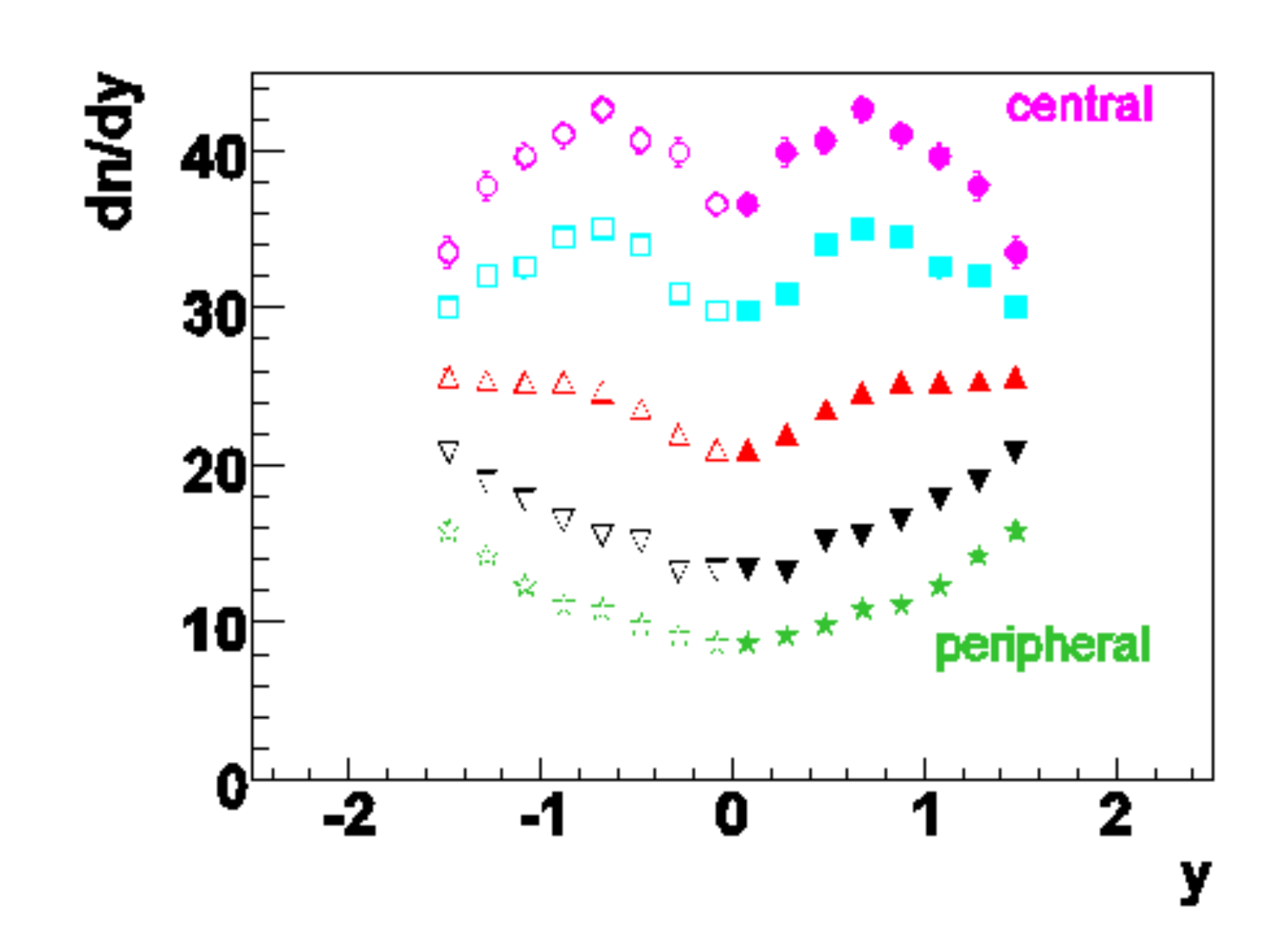}
\includegraphics[width=0.5\linewidth]{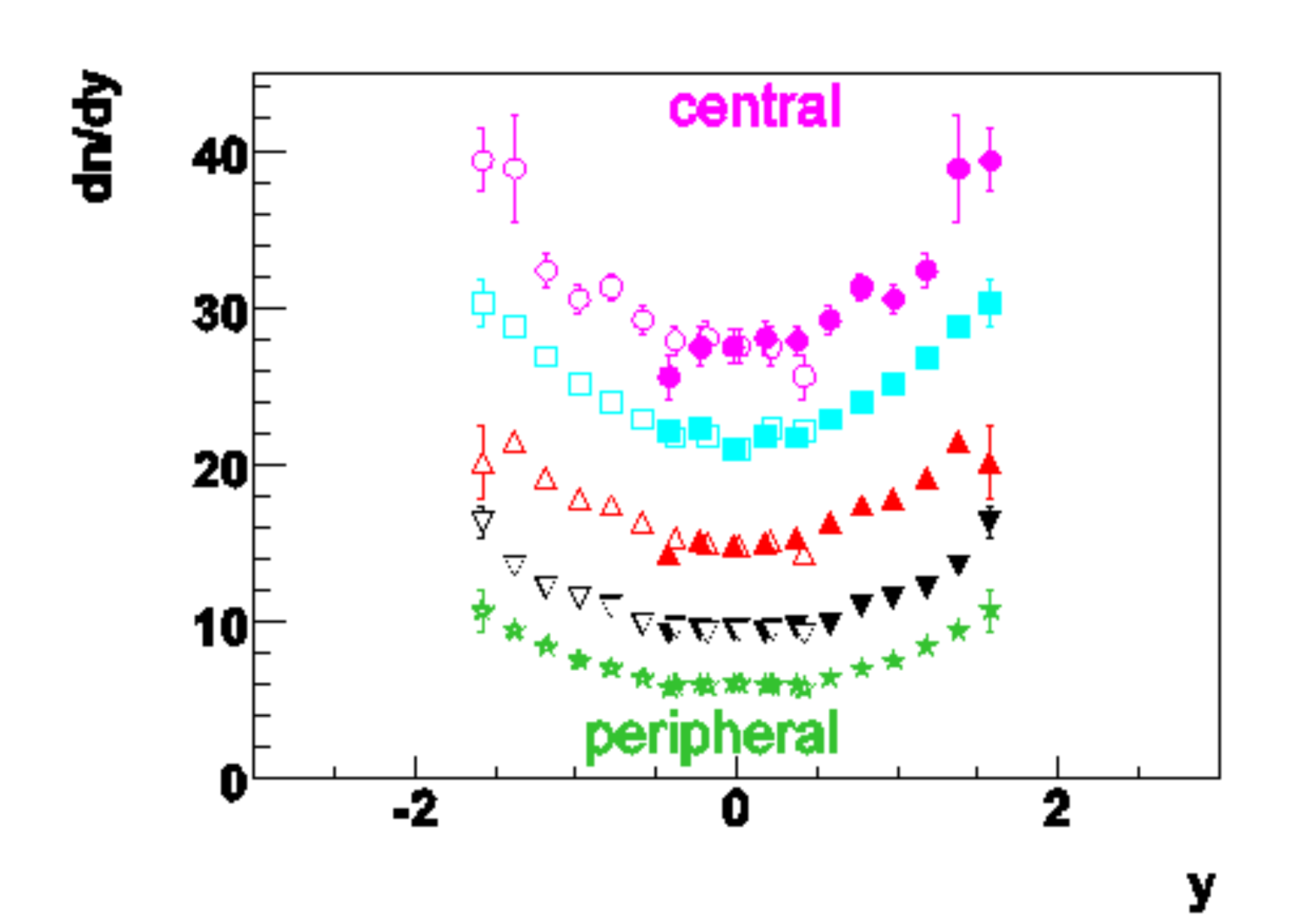}
\caption{
Proton rapidity distributions from Pb+Pb collisions at 40 (left) and 158 GeV/nucleon (right) for different centralities.The full symbols
stand for measured points.The open symbols are reflected at mid-rapidity. Beam rapidities are 2.2 at 40 and 2.9 at 158 
GeV/nucleon.
}
\label{fig:dndy_40_158}
\end{figure} 
\begin{figure}[h]
\includegraphics[width=0.85\linewidth]{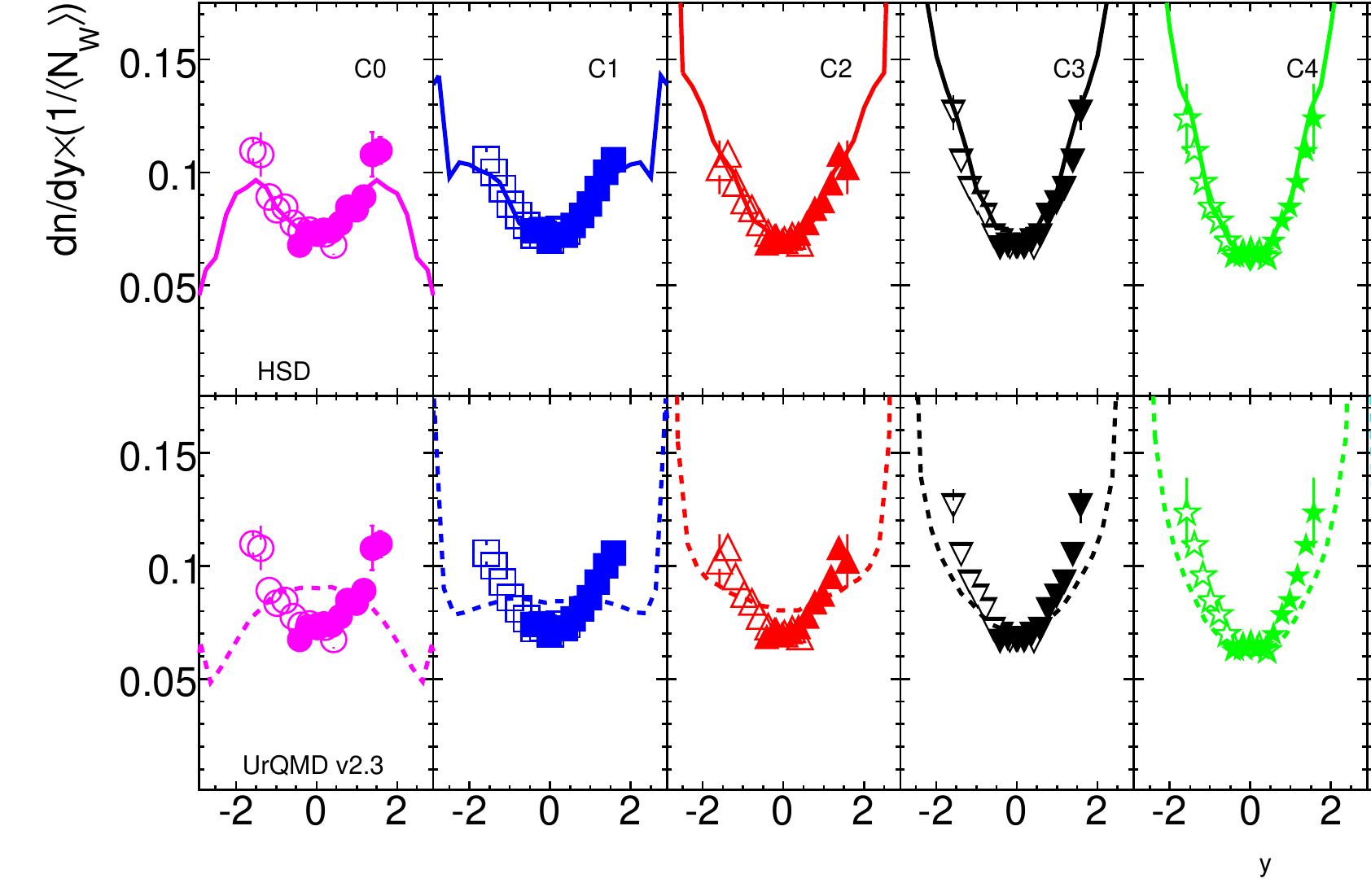}
\caption{
Proton rapidity distributions from Pb+Pb collisions at 158 GeV/nucleon for different centralities. The full symbols
stand for measured points. The open symbols are reflected at mid-rapidity. All spectra are normalized to the number of wounden nucleons (\nwound). Also shown are the results of model calculations
(HSD and UrQMD)
}
%\end{center}
\label{fig:dndy_158_modelcomp}
\end{figure} 
\end{center}
\Fi{fig:dndy_40_158} presents the main experimental results of our study. Proton rapidity distributions are shown at 5 different centralities and two beam energies. The centrality intervals range from the 0 - 5\%  most central events to the ensemble of events which populate the interval from 33.5 to 43.5\%. We note approximate shape independence at 158 GeV/nucleon (right). At  40 GeV/nucleon (left) the shape shows a similar evolution as at AGS (see \Fi{fig:AGS})  although a dip at mid-rapidity persists at all energies. We compare our finding at 158 GeV/nucleon with the microscopic transport models HSD \cite{Weber:2002qb} and
UrQMD \cite{Petersen:2008kb}
in \Fi{fig:dndy_158_modelcomp}  (model calculations at 40 GeV/nucleon results were not available at the time of the workshop). The HSD model does remarkably well in shape and in magnitude at all centralities. In UrQMD the protons are slowed down too much in central collisions and too little in the more peripheral ones. 

\begin{figure}[h]
\begin{minipage}[h]{7cm}
\includegraphics[width=1.0\textwidth]{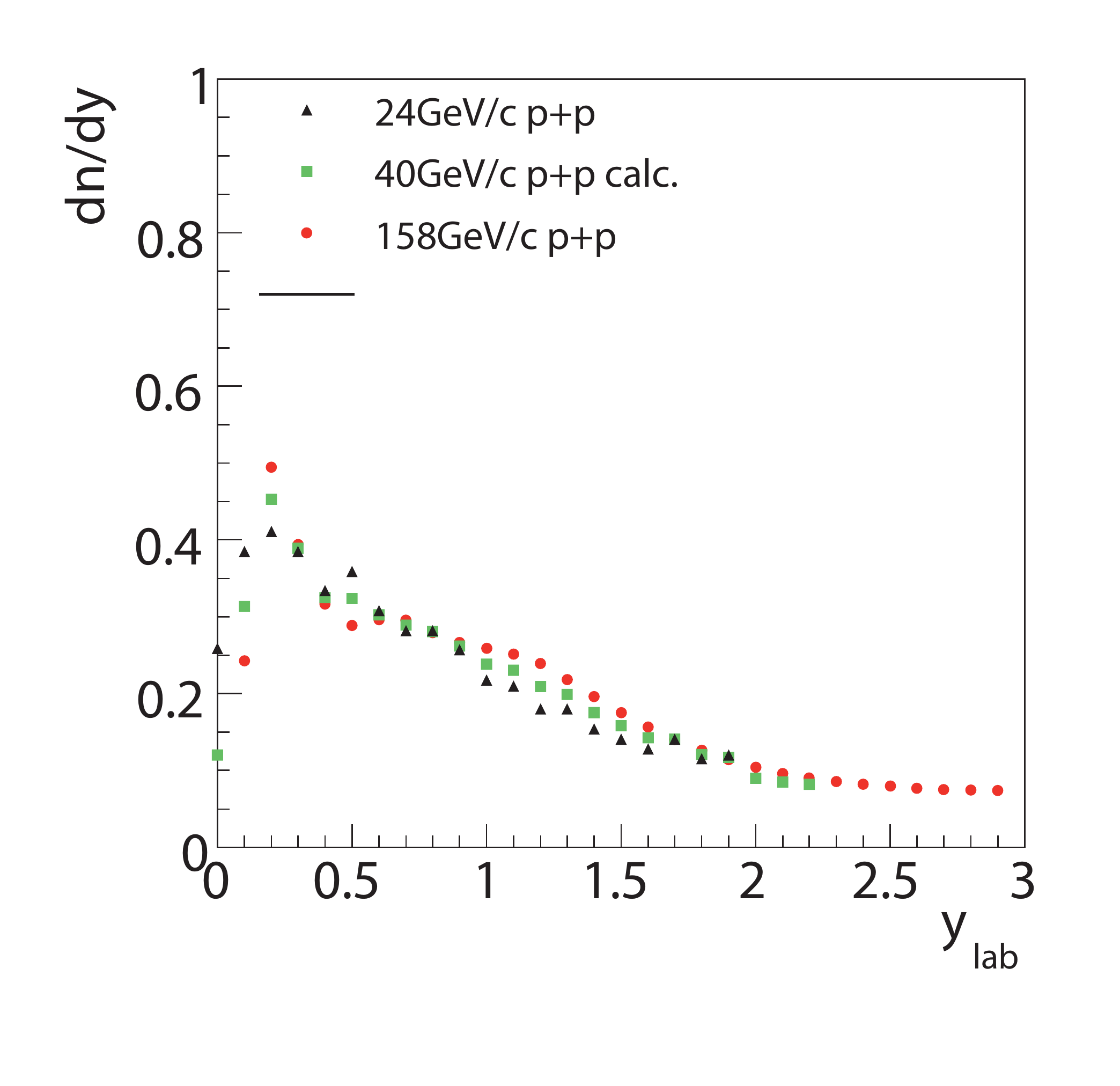}
\caption{
Proton number density as function of laboratory rapidity from inelastic p+p interactions at 158 GeV/c (red circles). The experimental data were measured in the forward hemisphere \protect{\cite{Fischer}}, but are shown here in the backward hemisphere (by reflection at mid-rapidity). Also shown are experimental results from p+p interactions at 24 GeV/c (triangles) \protect{\cite{Blobel}} and the average of the two (squares).
}
\label{fig:dndy_158_pp}
\end{minipage}
\hspace{0.9cm}
\begin{minipage}[h]{7cm}
\vspace{-1.3cm}
\includegraphics[width=1.0\textwidth]{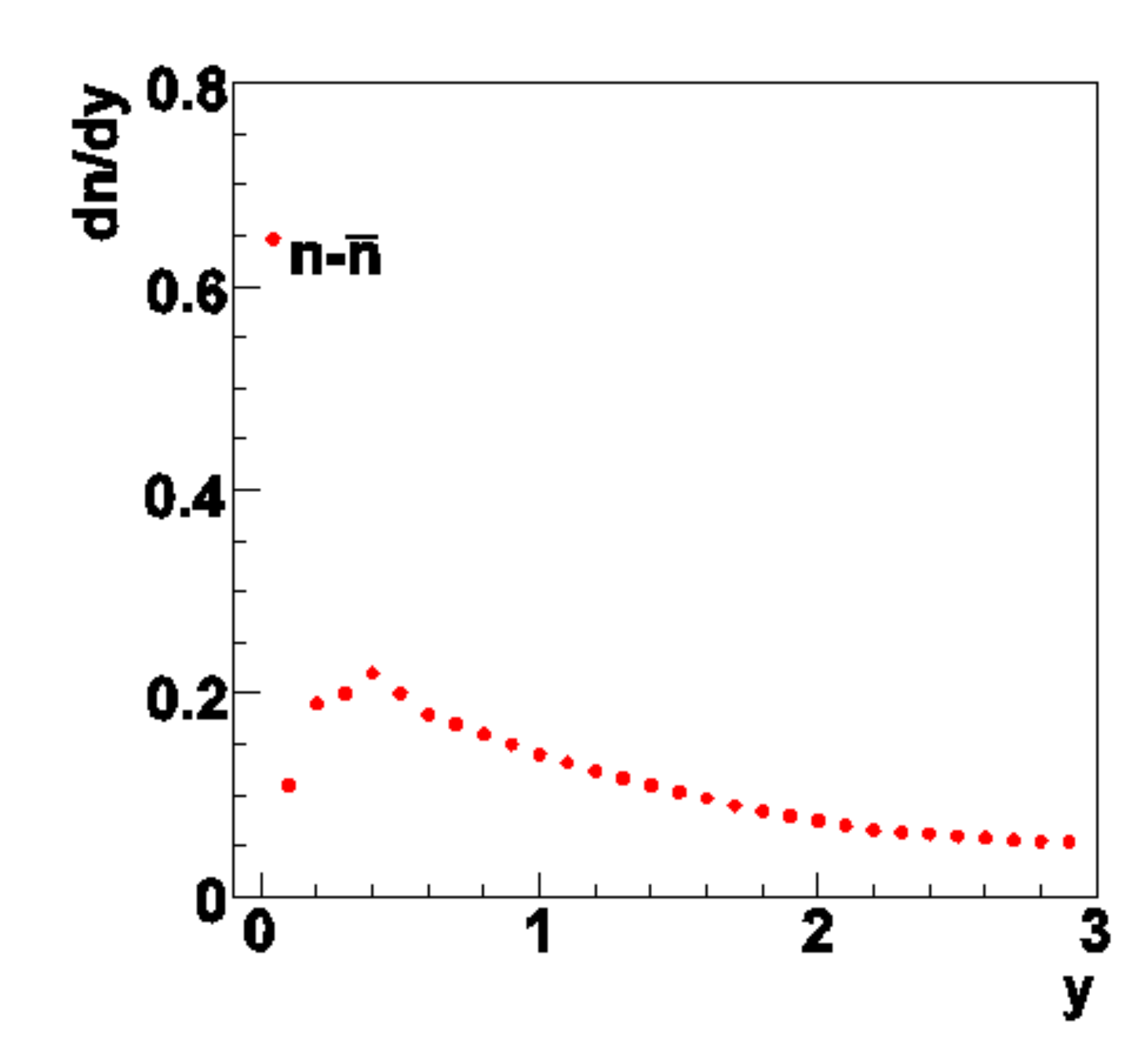}
\vspace{0.1cm}
\caption{
Net neutron number density as function of laboratory rapidity from inelastic p+p interactions at 158 GeV/c. The experimental data were measured in the forward hemisphere but are shown here in the backward hemisphere (obtained by reflection at mid-rapidity). 
}
\label{fig:dndy_neutron_pp}
\end{minipage}
\end{figure}

Next we compare the proton rapidity spectra of p+p and Pb+Pb both for experimental and model data. We first take a look at the new NA49 results on proton distributions in inelastic p+p collisions at 158 GeV/c \cite{Fischer}. \Fi{fig:dndy_158_pp} shows dn/dy of net protons\footnote{Net protons are obtained by subtracting the anti-protons from the protons and are the appropriate observable for the present study. However since the the $\overline{p}/p$ ratio is small at all rapidities at 40 GeV/nucleon the subtraction has not been done in this dataset. We have stated explicitely "net protons" whenever subtracted spectra are presented in the 158 GeV/nucleon dataset.} obtained by NA49 as function of laboratory rapidity together with 24 GeV/c data \cite{Blobel}. The shapes of both proton distributions are very similar. We can thus use their average as reference for the comparison of  the Pb+Pb data at 40  GeV/nucleon. However p+p and Pb+Pb do not have the same isospin composition. In fact the proper reference would be the appropriately averaged proton and neutron distributions from inelastic p+p, p+n, and n+n interactions. Since the latter two are not availble, we consider the sum of protons and neutrons from p+p as reasonable approximation. The new NA49 data on neutron production in p+p interactions at 158 GeV were measured as function of x$_F$ \cite{Fischer}. We have transformed this distribution into dn/dy using the appropriate Jacobian and for the mean transverse masse values between 1.1 and 0.97 GeV/c$^2$ at y$_{cm}$ = 0.8 and 2.8 respectively. The result is shown in \Fi{fig:dndy_neutron_pp}.  

\begin{figure}[h]
\begin{center}
\includegraphics[width=0.7\linewidth]{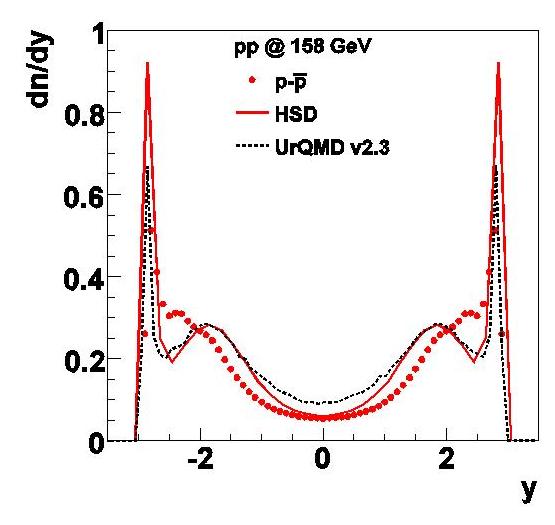}
\caption{
Net proton rapidity distributions from inelastic p+p interactions at 158 GeV/c \protect{\cite{Fischer}}. The full circles represent the experimental data
which were measured in the forward hemisphere and are reflected at mid-rapidity. Also shown are results of
calculations with the HSD (top) \protect{\cite{Weber:2002qb}} and UrQMD (bottom) \protect{\cite{Petersen:2008kb}} models.
}
\label{fig:dndy_158_pp_modelcomp}
\end{center}
\end{figure} 

Before we come to the detailed comparison of net proton rapidity distributions from Pb+Pb and p+p collisions we check how well the two models reproduce the new p+p data in \Fi{fig:dndy_158_pp_modelcomp}.
The overall features of the experimental data are reasonably well described by HSD and UrQMD. Significant differences emerge only next to the elastic peak at rapidities near to 2.5~. Here the models underpredict the experimental data by almost 30\%. This loss is compensated by an overshoot around rapidities of 1.5.

\begin{figure}[h]
\includegraphics[width=0.5\textwidth]{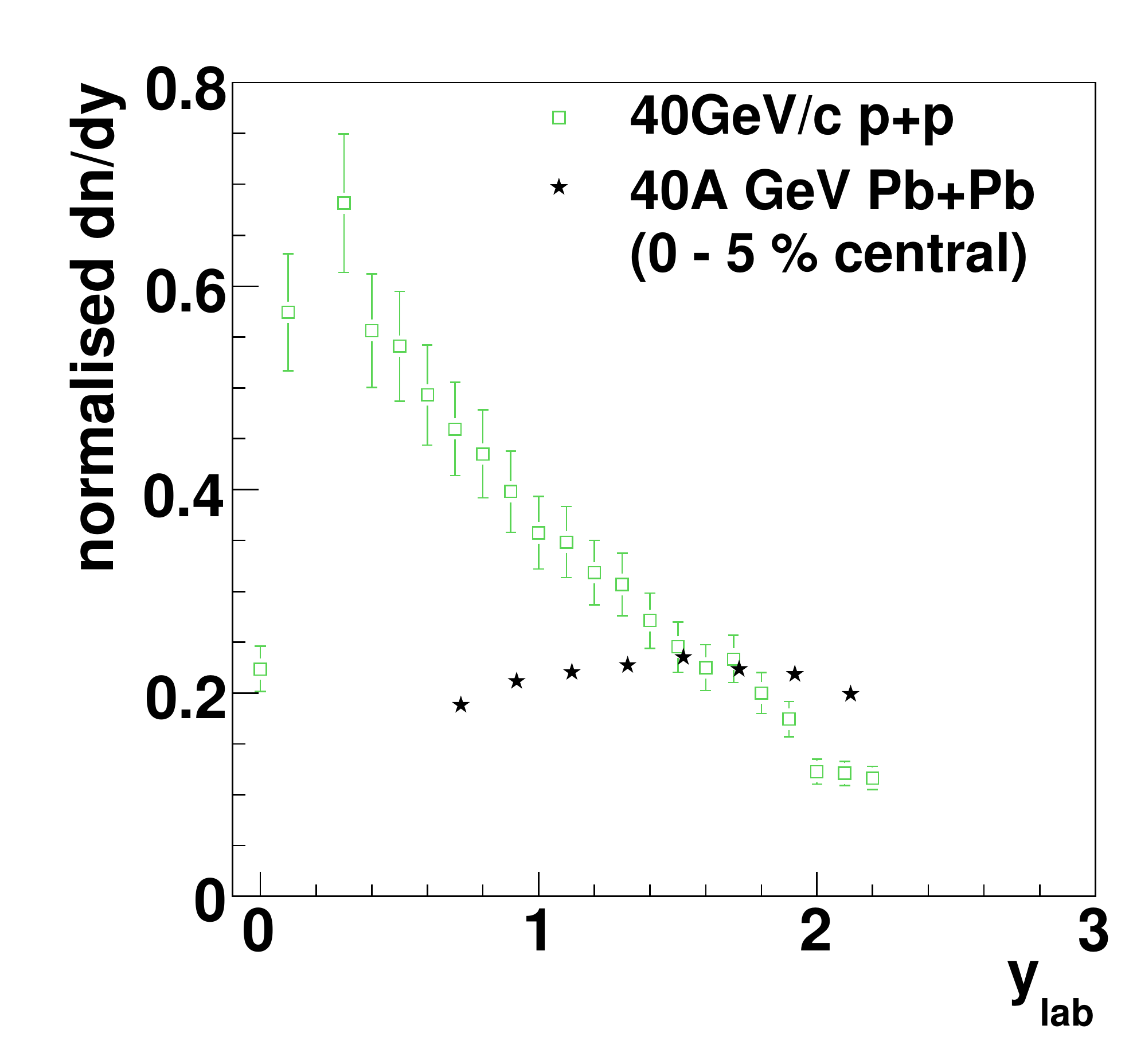}
\includegraphics[width=0.5\textwidth]{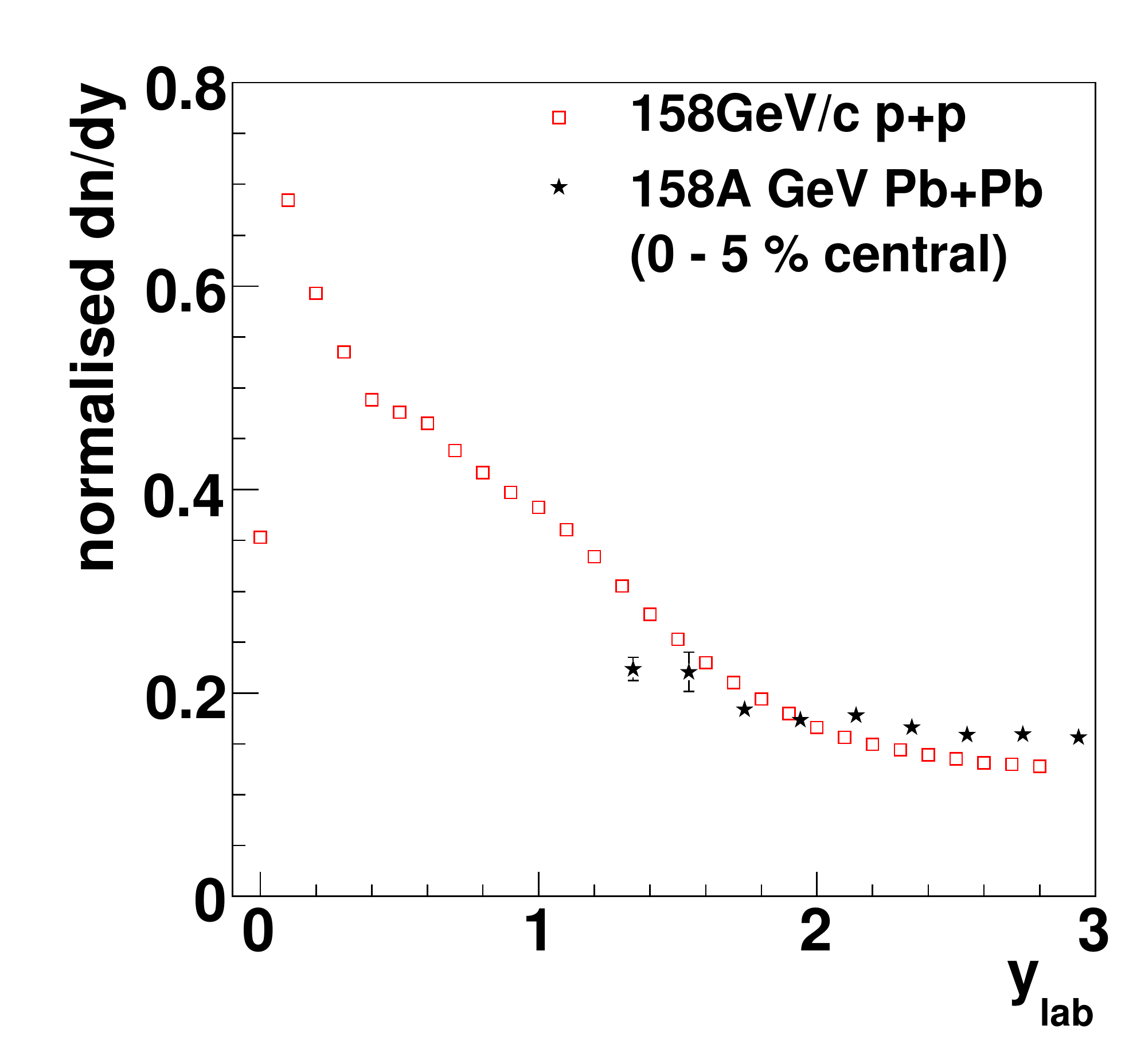}
\includegraphics[width=0.5\textwidth]{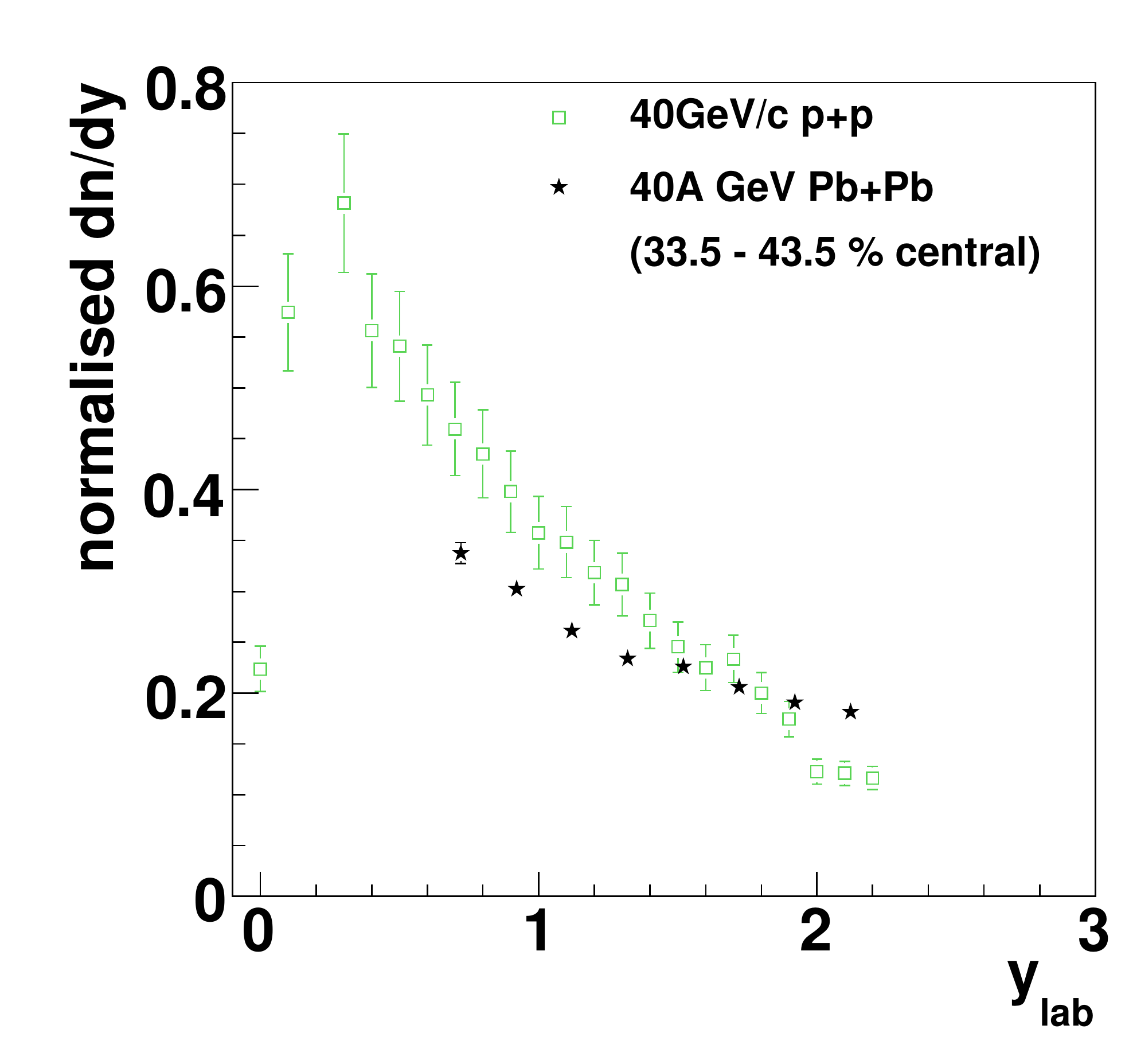}
\includegraphics[width=0.5\textwidth]{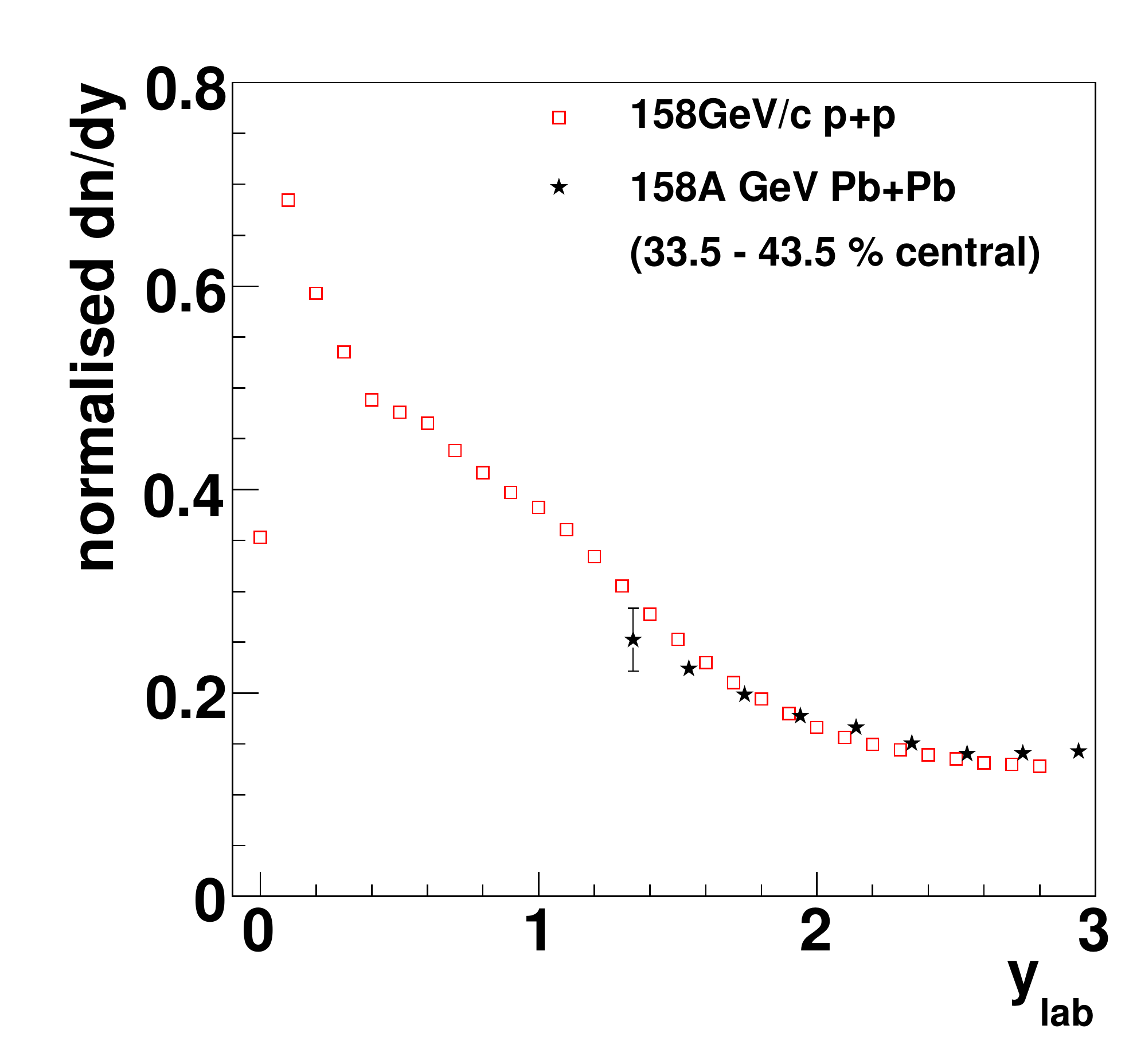}
\
\caption{
Normalized net proton rapidity distributions from Pb+Pb collisions at 40 (left) and 158 GeV/nucleon (right) for the most central 
(upper row) and the most peripheral centrality selection (lower row). The normalization constant is \nwound/2 to account for the unseen neutrons in the Pb+Pb data. Also shown as reference are the rapidity distributions of net protons plus neutrons from inelastic p+p interactions (open symbols) . 
}
\label{fig:comp_pp_Pb}
\end{figure} 

\Fi{fig:comp_pp_Pb} summarizes the most important conclusion from the study of proton production in Pb+Pb and p+p collisions at SPS energies. We compare the yield of net nucleons (protons plus neutrons) as function of rapidity, normalised to the average number of wounded nucleons from p+p interactions with net protons from Pb+Pb collisions at two centralities (most central in the upper and most peripheral in the lower row) and two energies (40 GeV/nucleon on the left and 158 GeV/nucleon on the right). The p+p points at the lower energy are from an interpolation of 158 and 24 GeV/c data (see \Fi{fig:dndy_158_pp}). We find that the Pb+Pb and p+p spectra are remarkably similar in shape and in yield. In particular there is almost perfect agreement between Pb+Pb and p+p data in peripheral collisions at the higher energy (lower right). In central collisions the parabolic shape gets more shallow but the integral of the normalized measured spectrum remains almost unchanged (0.39 in central and 0.41 in peripheral collisions). We note here that this integral over the complete rapidity interval would be two, if the net baryon content in the hyperons and in bound nucleons is neglected.
It seems that the effects due to secondary interactions of the participant nucleons happen mostly outside of the experimental acceptance at the higher energy. The multiple collision effects are coming out clearer at the lower energy (left column). At 40 GeV/nucleon a rapid, significant change in shape occurs roughly one unit away from mid-rapidity or 1.2 units away from target-rapidity. We note that this region (1.2 units away from target-rapidity) is not covered by our acceptance at the higher energy (158 GeV/nucleon). The effect observed at 40 GeV/nucleon could thus also be present at 158  GeV/nucleon. The integrated yields of the normalized 40 GeV/nucleon spectra are again remarkably constant (0.36 in central and 0.38 in peripheral collisions). The comparison to our interpolated p+p reference is less conclusive than at 158 GeV/nucleon. It is certainly desirable to get the corresponding experimental data from the forthcoming NA61 experiment (see contribution by G. Stefanek to these proceedings).

In view of the significant differences in the number of collisions per participating nucleon between p+p and Pb+Pb as well as in peripheral and central (Pb+Pb) collisions our findings suggest that multiple collisions have little influence on the protons close to mid-rapidity, i.e. those which experience strong stopping. 
This finding is more pronounced at 158 than at 40 GeV/nucleon. It could be verified by a detailed study of transport model caculations which seem to reproduce these phenomena.  

%\newpage

\end{document}